# Materials & Properties: Mechanical Behaviour

*C. Garion*
CERN, Geneva, Switzerland

**Abstract**
All systems are expected to be designed to fulfil their functions over their requested lifetime. Nevertheless, failure of a system may occur, and this is unfortunately, true also for vacuum systems. From a mechanical point of view, buckling, leak by fatigue crack propagation, rupture of a fixed support under unbalanced vacuum force… may happen. To well understand and anticipate the behaviour of a structure, it is mandatory to first understand the mechanical behaviour of materials as well as their failure modes. This is presented in a first part of the document. Then, the structural behaviour of a vacuum system and its relationship with the material properties is discussed. Finally, guidelines for the material selection for a given application are roughed out.

## 1 Introduction

Used as vacuum envelope, the material is of primary importance to ensure the mechanical reliability of the vacuum systems. In general, vacuum chambers have to be more and more optimized for mechanical or physics considerations, or to fulfil space requirements which are more and more stringent, or basically for economic reasons. To optimize the chambers and therefore, minimize margin to a reasonable value while ensuring the integrity of the system, the material properties, their behaviour and their influence on the mechanical behaviour of the structure have to be well understood.

The first chapter is devoted to different models of materials based on continuum mechanics. After simple linear isotropic elasticity, other notions from plasticity to fracture mechanics are presented. These models can be wisely used for applications where material failure might be expected (fatigue, initial defect propagation, etc.). The basic parameters of isotropic linear thermal elasticity to be used in most of the applications are introduced.

The second chapter deals with the design of the vacuum chambers as a structural component but also of the whole vacuum system as a mechanical system. A general approach is presented covering the main loads applied on a vacuum chamber. Two key points shall be kept in mind of all designers of vacuum systems: First, the stability of structures subjected to pressure and external pressure in particular for vacuum applications; second the pressure thrust force induced by bellows on the adjacent chambers and consequently on the supports. Despite these two phenomena are well known, numerous examples of failure with dramatic consequences can be found in accelerators.

The third chapter aims at giving some information about the material selection. This process uses the material properties but introduces also the figures of merit of materials for a given applications. The chapter is focused on the material selection based on their physical or mechanical properties. Other points, not covered in this paper, shall be considered as well for the material selection: the availability, the transformation process (machining, forming, etc.), the integration/joining to the system (welding or brazing), radiation compatibility (resistance and activation), cost, etc.

This paper is not intended to give an exhaustive view of materials and mechanics but aims at giving a comprehensive overview of both fields and how both are connected together and to the mechanical design of vacuum systems.

## 2 Material modelling

### 2.1 Basic notions

To characterize the mechanical properties and behaviour of materials, intrinsic variables are introduced. The stress in a section is defined as the ratio between the force applied on a section and its surface area. Usually denoted as $\sigma$, it is commonly expressed in MPa (N·mm$^{-2}$). Similarly, the strain is given by the relative variation of a given length. This variable is dimensionless.

Basic behaviour and mechanical properties are assessed from tensile tests carried out on a sample of probing length, L and section $S_0$. The force is measured as a function of the elongation. Then, the typical curve giving the nominal or engineering stress ($F/S_0$) as a function of the nominal strain ($\Delta L/L_0 = (L-L_0)/L$) is derived. Three main material properties are assessed: The Young's modulus, denoted E, corresponding to the stiffness of the material in the elastic domain; the yield stress $\sigma_y$ or $R_p$ (elastic limit) and the tensile strength $\sigma_m$ or $R_m$ (Fig. 1). The Poisson ratio, $\nu$, defined by the ratio between the transversal and axial strain can be determined as well by the tensile tests.

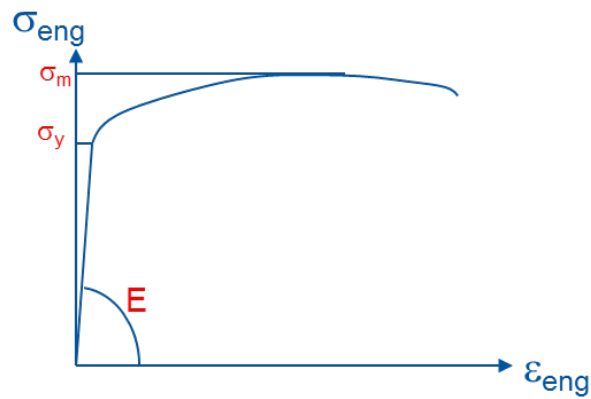

**Fig. 1:** Typical tensile curve of metals

Different categories of material behaviours can be observed. Among them:

- Elastic brittle material: the deformation is reversible, and failure occurs without noticeable overall plastic deformation (Fig. 2(a)).

- Elastic plastic behaviour: after elastic reversible deformation, slips occurs in crystals and lead to residual deformation after unloading (Fig. 2(b)).

- Elastic plastic damageable materials: for large deformation, microcracks and/or microvoids appear in the materials leading to noticeable decrease of the stiffness (Fig. 2(c)).

- Elastic viscoplastic behaviour: material response in the plastic domain depends on the strain rate (Fig. 2(d)).

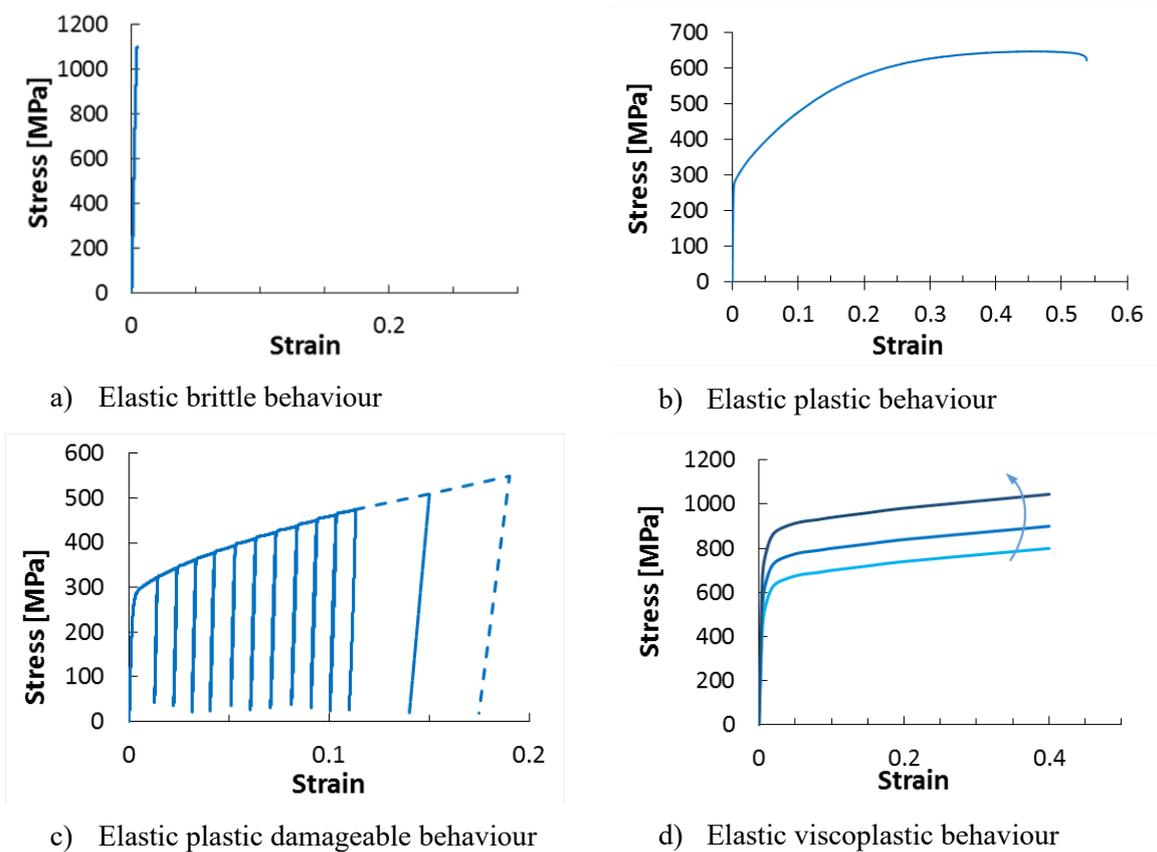

a) Elastic brittle behaviour  
b) Elastic plastic behaviour  
c) Elastic plastic damageable behaviour  
d) Elastic viscoplastic behaviour  

**Fig. 2:** Some basic material behaviours

The geometry of the sample evolves during the loading. Therefore, true variables, strain and stress, are introduced. They are defined from the engineering strain and stress by:

$$d\varepsilon = \frac{dl}{l} \rightarrow \varepsilon = \ln(1 + \varepsilon_{eng}) \tag{1}$$

$$\sigma = \frac{F}{S} \cong \sigma_{eng} \cdot (1 + \varepsilon_{eng}) \tag{2}$$

A comparison between the engineering stress-strain curve and the true stress-strain curve is shown in Fig. 3. Whereas the difference is not noticeable in the elastic domain, it has to be considered for a large strain to characterize correctly the materials.

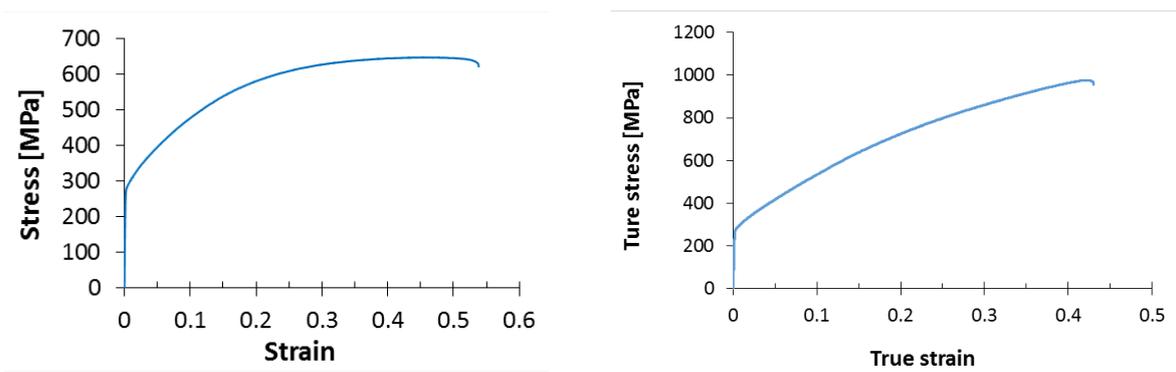

**Fig. 3:** Comparison of engineering and true variables

## 2.2 Notions of strain and stress in continuum mechanics

The stress vector **T** defined at a point **M** and for a given surface represented by its normal **n** is given by:

$$T(\boldsymbol{M}, \boldsymbol{n}) = \lim_{dS \to 0} \frac{d\boldsymbol{F}}{dS} \tag{3}$$

It can be written as a function of the stress tensor $\underline{\underline{\sigma}}$, defined by:

$$T(\boldsymbol{M}, \boldsymbol{n}) = \underline{\underline{\sigma}}(\boldsymbol{M}) \cdot \boldsymbol{n} \tag{4}$$

The normal and shear stress are called $\sigma_n$ and $\tau$.

They are defined by $\sigma_n = \boldsymbol{n} \cdot \underline{\underline{\sigma}}(\boldsymbol{M}) \cdot \boldsymbol{n}$ and $\tau = \underline{\underline{\sigma}}(\boldsymbol{M}) \cdot \boldsymbol{n} - \sigma_n \cdot \boldsymbol{n}$, respectively. In tension, $\sigma_n$ is positive. It can be shown that $\underline{\underline{\sigma}}$ is a symmetric tensor and is therefore represented in the principal base by a diagonal matrix with the principal stresses, $\sigma_I$, $\sigma_{II}$ and $\sigma_{III}$. The hydrostatic stress, $\sigma_h$, and deviatoric stress tensor $\underline{\underline{\sigma}}^D$ are defined by:

$$\underline{\underline{\sigma}} = \sigma_h \cdot \underline{\underline{I}} + \underline{\underline{\sigma}}^D \; ; \; \left(\sigma_h = \frac{1}{3} \cdot \mathrm{tr}\left(\underline{\underline{\sigma}}\right)\right) \tag{5}$$

From a kinematical point of view, considering the function $\Phi$, that transforms the initial configuration to the actual geometry ($\boldsymbol{M} = \Phi(\boldsymbol{M_0})$), the deformation gradient $\underline{\underline{F}}$ is defined by:

$$\mathrm{d}\boldsymbol{M} = \frac{\partial \Phi(\boldsymbol{M_0})}{\partial \boldsymbol{M_0}} \cdot \mathrm{d}\boldsymbol{M_0} = \underline{\underline{F}}(\boldsymbol{M_0}) \cdot \mathrm{d}\boldsymbol{M_0} \tag{6}$$

The Green Lagrange deformation is introduced by:

$$\underline{\underline{E}}(\boldsymbol{M_0}) = \frac{1}{2} \cdot \left[\underline{\underline{F}}^T(\boldsymbol{M_0}) \cdot \underline{\underline{F}}(\boldsymbol{M_0}) - \underline{\underline{I}}\right] \tag{7}$$

Considering the displacement ($\boldsymbol{M} = \boldsymbol{M_0} + \boldsymbol{u}(\boldsymbol{M_0})$) and small deformations, the strain tensor reads:

$$\underline{\underline{\varepsilon}}(\boldsymbol{u}) = \frac{1}{2} \cdot \left[\frac{\partial \boldsymbol{u}(\boldsymbol{M_0})}{\partial \boldsymbol{M_0}} + \left(\frac{\partial \boldsymbol{u}(\boldsymbol{M_0})}{\partial \boldsymbol{M_0}}\right)^T\right] \tag{8}$$

Finally, the energy variation in the material of a body $\Omega$ is: $\mathrm{d}\mathbb{E} = \oiiint_\Omega \underline{\underline{\sigma}} : \mathrm{d}\underline{\underline{\varepsilon}}$.

## 2.3 Thermal elasticity

### *2.3.1 Linear elasticity*

All materials are composed of atoms which are held together by bonds resulting from electromagnetic interactions. When external load is applied, the lattice is deformed, interatomic distances change leading to interatomic forces that, in a first approach, are proportional to the interatomic distance variation. This deformation is reversible. This behaviour at the atomic level can be extended at the mesoscopic scale under the Hooke's law:

$$\underline{\underline{\sigma}} = \underline{\underline{C}} : \underline{\underline{\varepsilon}}^e \text{ with } \underline{\underline{C}} \text{ the stiffness or elastic tensor} \tag{9}$$

For homogeneous and isotropic materials, the elastic tensor is defined by only two independent parameters:

$$\underline{\underline{\sigma}} = \lambda \cdot \mathrm{tr}\left[\underline{\underline{\varepsilon}}^e\right] \cdot \underline{\underline{I}} + 2 \cdot \mu \cdot \underline{\underline{\varepsilon}}^e \text{ or } \underline{\underline{\varepsilon}}^e = \frac{1+\nu}{E} \cdot \underline{\underline{\sigma}} - \frac{\nu}{E} \cdot \mathrm{tr}\left[\underline{\underline{\sigma}}\right] \cdot \underline{\underline{I}} \tag{10}$$

$\lambda$ and $\mu$ stand for the Lamé coefficients. $\mu$, sometimes denoted $G$, is the shear modulus. The relationships between the elastic parameters are given in the set:

$$\lambda = \frac{\nu E}{(1+\nu)(1-2\nu)} \text{ and } \mu = \frac{E}{2(1+\nu)}$$
$$E = \mu \frac{3\lambda+2\mu}{(\lambda+\mu)} \text{ and } \nu = \frac{\lambda}{2(\lambda+\mu)} \quad (11)$$
$$k = \lambda + \frac{2\mu}{3} = \frac{E}{3(1-2\nu)}$$

The Young's modulus is a direct consequence of the type of bonds. Typical values are given in Tables 1 and 2 for different bonds and metals, respectively.

**Table 1:** Typical range for Young's modulus

| Bond type | Typical range of the Young's modulus [GPa] |
|---|---|
| Covalent | 1000 |
| Ionic | 50 |
| Metallic | 30-400 |
| Van der Waals | 2 |

**Table 2:** Elastic parameters at room temperature for metals commonly used for UHV applications

|  | Aluminium | Copper | Stainless steel | Titanium |
|---|---|---|---|---|
| E [GPa] | 70 | 130 | 195 | 110 |
| $\nu$ | 0.34 | 0.34 | 0.3 | 0.32 |

### 2.3.2 *Thermal linear isotropic elasticity*

A temperature variation changes the equilibrium state between atoms and introduces a free thermal strain. It is represented by the thermal strain tensor, $\underline{\underline{\varepsilon}}^{th}$, that is defined for an isotropic material by:

$$d\underline{\underline{\varepsilon}}^{th} = \alpha(T) \cdot dT \cdot \underline{\underline{I}} \quad (12)$$

with $\alpha$ the coefficient of thermal expansion (CTE).

For temperature independent dilatation coefficient, the thermal strain reads with respect to a temperature of reference $T_{ref}$:

$$\underline{\underline{\varepsilon}}^{th} = \alpha \cdot (T - T_{ref}) \cdot \underline{\underline{I}} \quad (13)$$

The total strain tensor is the sum of the thermal and mechanical contributions, $\underline{\underline{\varepsilon}} = \underline{\underline{\varepsilon}}^e + \underline{\underline{\varepsilon}}^{th}$, leading to:

$$\underline{\underline{\sigma}} = \underline{\underline{\underline{C}}} : \underline{\underline{\varepsilon}}^e = \underline{\underline{\sigma}} = \underline{\underline{\underline{C}}} : \left(\underline{\underline{\varepsilon}} - \underline{\underline{\varepsilon}}^{th}\right) \quad (14)$$

In one dimensional case, it reads simply:

$$\sigma = E \cdot \left(\varepsilon - \varepsilon^{th}\right) = E \cdot (\varepsilon - \alpha \cdot \Delta T) \quad (15)$$

Therefore, for a free stress state, the strain is $\alpha \cdot \Delta T$ and consequently the dilatation of a structure of length $L$ is $\alpha \cdot \Delta T \cdot L$. For a rigidly clamped structure, the strain is null leading to thermal stress equal to $\sigma = -E \cdot \alpha \cdot \Delta T$. In this case, the thermal stresses and consequently the forces generated by high temperature variation are high and shall be avoided. That's why only one fixed support shall be used along a baked vacuum chamber.

The evolution of the thermal contraction as a function of temperature is given in Fig. 4 for some materials as well as the CTE at room temperature in Table 3.

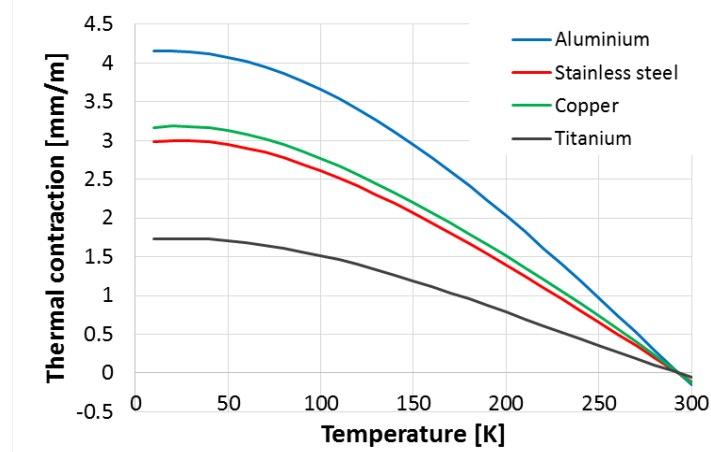

**Fig. 4:** Thermal contraction of some materials

**Table 3:** Coefficient of thermal expansion at room temperature for metals commonly used for UHV applications

|  | **Aluminium** | **Copper** | **Stainless steel** | **Titanium** |
|---|---|---|---|---|
| $\alpha$ [$10^{-6} \cdot K^{-1}$] | 22 | 17 | 16 | 8.9 |

## 2.4 Strength of materials

After elastic deformations, two main material behaviours are observed:
- A brittle failure occurs without noticeable permanent deformation.
- A plastic deformation occurs before ductile failure.

### 2.4.1 *Strength of brittle materials*

When local strain energy due to external loads becomes equal to the energy needed to pull the atom layers apart, a brittle failure by cleavage occurs. This failure is enhanced at low temperature and can be inter or intragranular. The material is very sensitive to stress concentration and the strength of brittle materials strongly depends on initial defects. Therefore, a scattering of the material strength is observed and is usually represented by the Weibull's law (Eq. (16)). It defined the survival probability of a given volume $\Omega$ under an equivalent stress level.

$$P_S^\Omega = exp\left(-\frac{1}{V_0}\int_\Omega \left(\frac{\sigma_{eq}}{\sigma_0}\right)^m d\Omega\right) \quad (16)$$

$m$ and $\sigma_0$ are the two Weibull's coefficients called the shape and scale parameters, respectively. The equivalent stress is the maximal positive principal stress.

## 2.4.2 Plasticity

For materials with plasticity, irreversible plastic deformation occurs before rupture. The mechanism is based on the dislocation motions and slip deformations under shear stress. The change of volume due to the dislocation motions or density is negligible (change of volume is due to interatomic distance and therefore thermoelastic deformation). The yield stress leading to inelastic deformation is not easily identifiable. Conventionally, the 0.2 % and 1 % proof strength are introduced considering the stress leading to 0.2 and 1 % of plastic strain (Fig. 5).

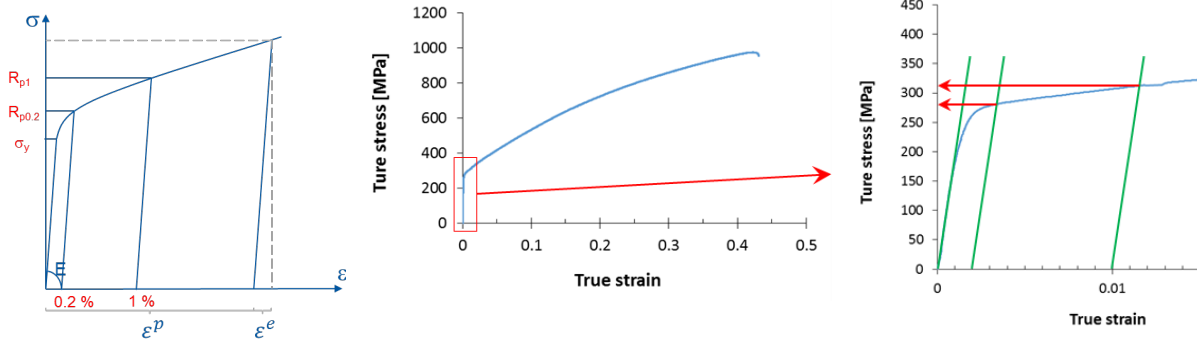

**Fig. 5:** Definition and identification of yield and proof strength

The total strain is the sum of the elastic and plastic strain. The elasticity condition is given by $\sigma < \sigma_y$ ($\sigma - \sigma_y < 0$). In practice, the yield strength is usually considered equal to the 0.2% proof strength: $\sigma_y = R_{p0.2}$.

### 2.4.2.1 Yield surface

In the theory of plasticity, the elasticity domain is defined by a yield surface, $f$, function of the stress, material strength and that verifies:

$$f\left(\underline{\underline{\sigma}}, \sigma_y, \dots\right) < 0 \text{ for elasticity}$$
$$f\left(\underline{\underline{\sigma}}, \sigma_y, \dots\right) = 0 \text{ for plasticity} \quad (17)$$

Two isotropic criteria are mainly used for isotropic metals:

- Tresca criterion: Based on the maximum shear stress, that is expressed as of function of the principal stresses by $\tau_{max} = Max(\sigma_i - \sigma_j)$, the Tresca's yield surface reads:

$$f\left(\underline{\underline{\sigma}}, \sigma_y\right) = Max(\sigma_i - \sigma_j) - \sigma_y \quad (18)$$

- Von Mises criterion: Based on the distortional elastic energy, proportional to $\underline{\underline{\sigma}}^D : \underline{\underline{\sigma}}^D = \text{tr}\left[\underline{\underline{\sigma}}^D \cdot \underline{\underline{\sigma}}^D\right]$, the Von Mises' yield surface reads:

$$f\left(\underline{\underline{\sigma}}, \sigma_y\right) = \sqrt{\frac{3}{2} \cdot \underline{\underline{\sigma}}^D : \underline{\underline{\sigma}}^D} - \sigma_y \quad (19)$$

Equivalent Von Mises stress is given by:

$$\sigma_{VM} = \sqrt{\frac{3}{2} \cdot \underline{\underline{\sigma}}^D : \underline{\underline{\sigma}}^D} \quad (20)$$

In principal stress space and stress space it reads respectively:

$$f\left(\underline{\underline{\sigma}}, \sigma_y\right) = \frac{1}{\sqrt{2}} \cdot \sqrt{(\sigma_I - \sigma_{II})^2 + (\sigma_I - \sigma_{III})^2 + (\sigma_{II} - \sigma_{III})^2} - \sigma_y \quad (21)$$

$$f\left(\underline{\underline{\sigma}}, \sigma_y\right) = \frac{1}{\sqrt{2}} \cdot \sqrt{(\sigma_{11} - \sigma_{22})^2 + (\sigma_{11} - \sigma_{33})^2 + (\sigma_{22} - \sigma_{33})^2 + 6 \cdot (\sigma_{12}^2 + \sigma_{13}^2 + \sigma_{23}^2)} - \sigma_y$$

These two criteria are plotted in Fig. 6 for two plane stress state conditions. It turns out that Tresca's condition is slightly more conservative than Von Mises' criterion.

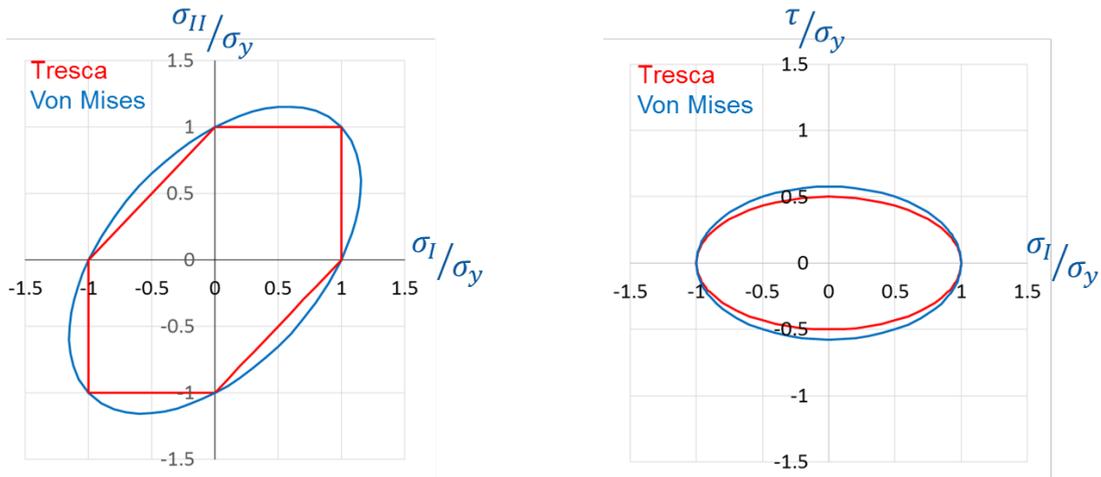

**Fig. 6:** Comparison of Von Mises and Tresca criteria for plane stress state

### 2.4.2.2 *Plastic flow and hardening*

The plastic strains are represented by a tensor defined by: $\underline{\underline{\varepsilon}}^p = \underline{\underline{\varepsilon}} - \underline{\underline{\varepsilon}}^e - \underline{\underline{\varepsilon}}^{th}$. Plastic strains are induced by dislocation motions without volume change. Thus, $\text{tr}\left[\underline{\underline{\varepsilon}}^p\right] = 0$. When plasticity occurs, the yield criterion *f=0* shall be verified. This condition translates in a modification of the yield surface. Fig. 7 represents simple evolutions of the yield surface for a tensile test.

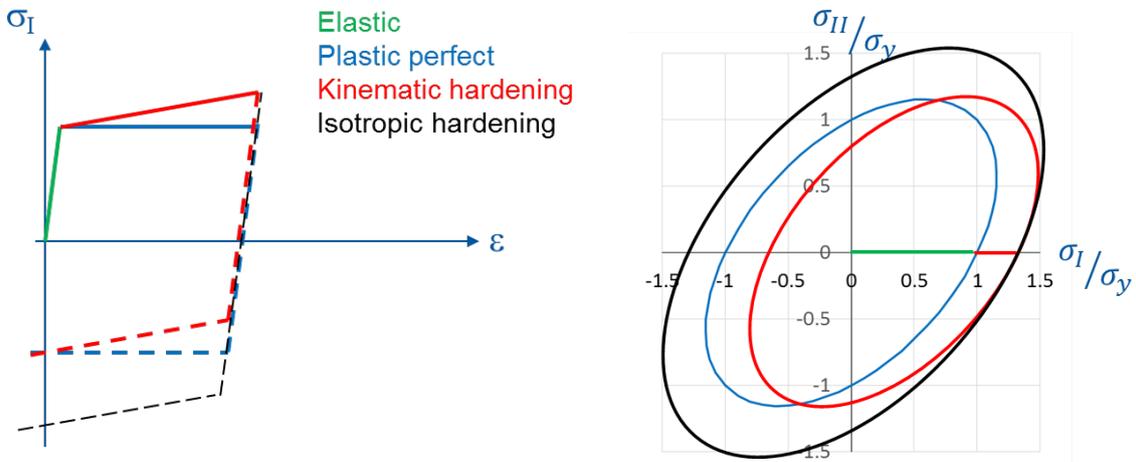

**Fig. 7:** Evolution of the plastic yield surface for different hardening cases

- Green line corresponds to the elastic behaviour until reaching the yield stress.
- For elastic perfect plastic behaviour, the stress is constant and equal to the yield strength. The yield surface does not change during loading.
- Red line corresponds to a kinematic hardening: the yield surface keeps the same shape but is displaced in the stress domain. For a reverse loading (compression), plasticity occurs at a lower stress level. This is called Bauschinger's effect. The kinematic hardening is associated roughly to the dislocation motions.
- In black, the yield surface is still centred with the stress origin, but the size increases during the loading (homothety with respect to the origin). For reverse loading, the plasticity is initiated for the same strength reached during loading. The isotropic hardening is associated to the dislocation density.

For real materials, a combination of the kinematic and isotropic hardening is usually observed. The hardening requires the introduction of additional internal variables for which evolution laws have to be defined. It is worth to mention that the difference of models may have only an impact for non-monotonic loading and that the yield surface depends on the loading path.

## 2.5 Failure of materials

The failure of materials or structure is a process, more or less long which starts with the apparition of microcracks or microvoids, followed by their growth and coalescence, leading finally to the initiation of a crack. This initial crack has a small size compared to the structure scale. It propagates until reaching the macroscopic rupture. Damage and fracture mechanics are used to model these two behaviours.

### 2.5.1 Damage mechanics

Progressive degradation of the materials is usually observed before the rupture. It corresponds to the material damage due to microvoids and/or microcracks. The phenomena leading to such defects are the motion and accumulation of dislocations, the decohesion or cleavage of precipitates or inclusions, intergranular decohesion, corrosion, irradiation, etc. Defects with size significantly smaller than the representative element scale, can be modelled by a damage variable defined a continuum damaged element. The damage variable is presented by a scalar for isotropic material, $D$ and is defined as the ratio between the surface of defects and the initial surface area. For a virgin structure, $D = 0$ whereas $D = 1$ means a crack initiation. An effective strain tensor is introduced.

$$\underline{\underline{\sigma}}' = \frac{\underline{\underline{\sigma}}}{1-D} \tag{22}$$

Assuming the strain equivalence principle, the constitutive law reads:

$$\underline{\underline{\sigma}} = \underline{\underline{C}}(1-D) : \underline{\underline{\varepsilon}}^e \tag{23}$$

This leads to the decrease of the apparent elasticity modulus: $E' = E \cdot (1-D)$. This property is commonly used to assess the initiation and evolution of damage in materials by applying loading/unloading during a tensile test (Fig. 2(c)).

For ductile materials, the damage evolution is related as a driving force to the plasticity and in particular to the accumulated plastic strain. The intensity of the damage evolution is linked to the strain energy release rate that can be expressed as a function of the effective Von Mises equivalent stress and the triaxiality ratio ($\sigma_h/\sigma_{VM}$). Damage mechanics can be extended to the assessment of the fatigue lifetime, $N_f$, as a function of the loads:

$$\Delta\varepsilon = \Delta\varepsilon^e + \Delta\varepsilon^p = \frac{C_2}{E} \cdot N_f^{-1/\gamma_2} + C_1 \cdot N_f^{-1/\gamma_1} \qquad (24)$$

For standard materials, it can be approximated by the universal slope equation:

$$\Delta\varepsilon = 3.5 \cdot \frac{\sigma_m}{E} \cdot N_f^{-0.12} + D_u^{0.6} \cdot N_f^{-0.6} \qquad (25)$$

where $D_u$ stands for a material parameter related to its ductility.

Equation leads to the Manson-Coffin's law used in the low cycle fatigue regime:

$$N_f \cdot (\Delta\varepsilon^p)^\beta = Cst \qquad (26)$$

### 2.5.2  Fracture mechanics

In presence of a crack in a body subjected to a stress $\sigma_\infty$, a stress concentration occurs at the crack tip. For elastic materials, the stress singularity reads:

$$\sigma \propto \frac{K(\sigma_\infty, a)}{\sqrt{2\pi \cdot r}} \qquad (27)$$

with the crack length and r, the distance from the crack tip. K denotes the stress intensity factor and reads in the form:

$$K(\sigma_\infty, a) = f \cdot \sigma_\infty \cdot \sqrt{\pi \cdot a} \qquad (28)$$

with $f$ a parameter depending of the crack and structure geometries. The $f$ parameter can be found in literature or estimate by numerical methods. Two examples are given in Table 4. Several solicitation modes of cracks are defined: mode I: Opening, mode II: in plane shear, mode III: out-of plane shear.

**Table 4:** Stress intensity factor for two simple cases

| Geometry | Stress intensity factor |
|---|---|
| 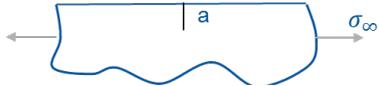 | $K_I = 1.122 \cdot \sigma_\infty \sqrt{\pi a}$ |
| 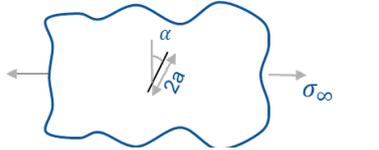 | $K_I = \sigma_\infty \sqrt{\pi a} \cos^2 \alpha$ <br> $K_{II} = \sigma_\infty \sqrt{\pi a} \cos \alpha \sin \alpha$ |

For brittle materials, brutal rupture occurs when the stress intensity factor is equal the material toughness. More general approaches based on energy considerations, can be used, in particular for ductile materials. Griffith's approach is based on the energy release rate defined by the elastic energy release during the crack propagation.

Fatigue crack growth rate curves exhibit typically a stable crack propagation phase followed by a sharp increase up to the critical toughness (Fig. 8).

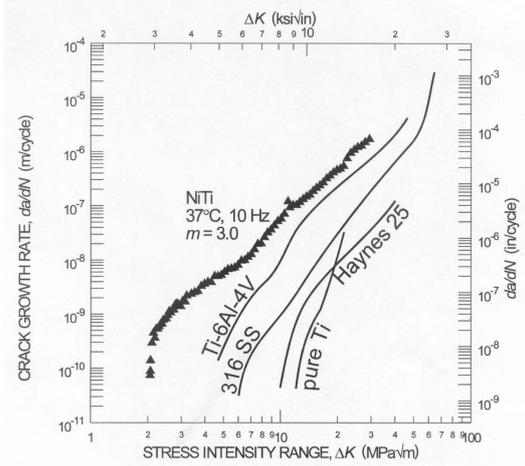 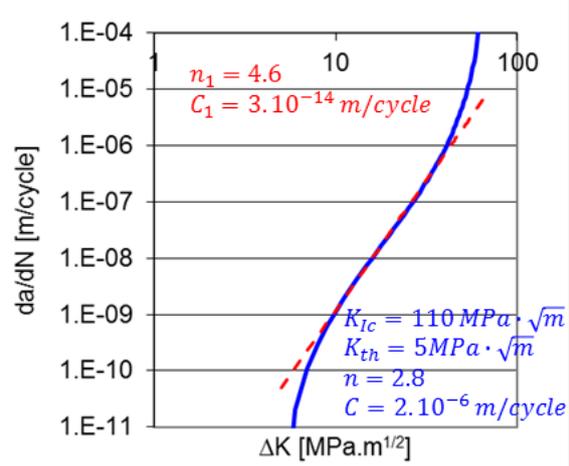

**Fig. 8:** Typical crack growth rate curves [1]   **Fig. 9:** Extrapolated crack propagation curves for 316

Paris' law is commonly used to assess the stable crack propagation in fatigue as a function of the stress intensity range.

$$\frac{da}{dN} = C_1 \cdot \Delta K^{n_1} \quad (29)$$

A more general rule covering the different phases can be used:

$$\frac{da}{dN} = C \left( \frac{K_{max} \frac{1-R}{1-mR} - K_{th}}{K_{Ic} - K_{max}} \right)^n \quad (30)$$

$K_{Ic}$ and $K_{th}$ are two fracture parameters. $R$ stands for the ratio $K_{min}/K_{max}$, $m$ denotes the average load parameter usually 0.5. A comparison of the two laws is given in Fig. 9 for 316 stainless steel.
It is worth to note that for some materials the fracture toughness is strongly temperature dependent and exhibit a ductile to brittle transition at low temperatures.

## 3  Structural mechanical analysis

### 3.1  Vacuum chamber

Vacuum chambers and beam pipes are designed according to sound engineering practice and generally in accordance with harmonised standard EN 13445 dedicated to unfired pressure vessel but also suggested for vacuum vessels. Specific vacuum chambers have to be optimised and a good knowledge of the design methodology is recommended.

#### *3.1.1  Loads on a vacuum chamber*

##### *3.1.1.1  Gravity*

For some applications in particular slender vacuum chamber, self-weight forces induced by gravity have to be considered. The specific forces read:

$$\boldsymbol{f_v} = \rho \cdot \mathbf{g} \quad (31)$$

$\rho$ and $\boldsymbol{g}$ stand for the specific mass and the gravitational acceleration, respectively.

The contribution of the pump or bake out jacket may be considered as well.

*3.1.1.2 Vacuum/pressure*

The force applied by a fluid with a pressure p on a surface dS of normal n reads simply:

$$\mathrm{d}\boldsymbol{F} = -p \cdot \mathbf{n} \cdot \mathrm{d}S \tag{32}$$

The continuity of the stress vector across the surface leads to:

$$\underline{\underline{\sigma}} \cdot \boldsymbol{n} = -p \cdot \boldsymbol{n} \tag{33}$$

For a uniform pressure, $\oiint p\boldsymbol{n}ds = 0$.

For a simple tube of radius $R$ and thickness $t$, the hoop (circumferential) stress and longitudinal stresses read respectively

$$\begin{aligned}\sigma_{\theta\theta} &= \frac{p \cdot R}{t} \\ \sigma_{zz} &= \frac{p \cdot R}{2 \cdot t}\end{aligned} \tag{34}$$

*3.1.1.3 Electromagnetic forces*

The variation of magnetic fields induces eddy currents (Foucault currents) in a conductive material. This effect is of primary importance for fast magnets or superconducting magnets during a quench (a few kA eddy currents during a LHC magnet quench). It is driven by the Maxwell's equation, given by:

$$\boldsymbol{rot\ E} = -\frac{\partial \boldsymbol{B}}{\partial t} \tag{35}$$

with $E$ the electrical field and $B$ the magnetic field. The current density is given by the local Ohm's law:

$$\boldsymbol{j} = \frac{\boldsymbol{E}}{\rho_e} \tag{36}$$

with $\rho_e$ the electrical resistivity. For long and simple structures (vacuum chambers, beam screens), the current direction is in axis (z direction). The coexistence of the eddy currents and the magnetic field generates Lorentz's or Laplace's forces on the conductor:

$$\boldsymbol{f} = \boldsymbol{j} \times \boldsymbol{B} \tag{37}$$

For a dipole and quadrupole magnetic field, local Lorentz's forces are derived, and integrated value are obtained by considering a circular vacuum chamber in **z** direction.

Table 5: Magnetic forces on a circular vacuum chamber

| Magnet type | Dipole | quadrupole |
|---|---|---|
| Magnetic field | $\boldsymbol{B} = B\mathbf{y}$ | $\boldsymbol{B} = G \cdot r \cdot \begin{vmatrix} \sin\varphi \\ \cos\varphi \\ 0 \end{vmatrix}_{(x,y,z)}$ |
| Specific Lorentz force | $\boldsymbol{f} = B\dot{B}\dfrac{x}{\rho_e}\mathbf{x}$ | $\boldsymbol{f} = \dfrac{1}{2\rho_e}G\dot{G}r^3\cos 2\varphi \begin{vmatrix} -\cos\varphi \\ \sin\varphi \\ 0 \end{vmatrix}_{(x,y,z)}$ |
| Integrated Lorentz force (per half or quadrant) | $\boldsymbol{F} = -2tB\dot{B}\dfrac{R^2}{\rho_e}\mathbf{x}$ | $\boldsymbol{F} = -\dfrac{\sqrt{2}}{3}tG\dot{G}\dfrac{R^4}{\rho_e}\mathbf{x}\,(\mathbf{y})$ |
| Integrated current (per half or quadrant) | $\boldsymbol{I} = -2t\dot{B}\dfrac{R^2}{\rho_e}\mathbf{z}$ | $\boldsymbol{I} = t\dot{G}\dfrac{R^3}{2\rho_e}\mathbf{z}$ |

To give some orders of magnitude, a copper circular vacuum chamber, 0.1 mm thick (coating) and 25 mm in radius, is considered. A dipole magnetic field and decay of 10 T and 100 T·s$^{-1}$, respectively, are assumed as well as an electrical resistivity of 10$^{-9}$ Ω·m. The integrated current in half of the chamber is 12.5 kA and the integrated Lorentz's force is 125 N·mm$^{-1}$ (12.5 tons per vacuum chamber meter).

It is worth to mention that the electrical resistivity is strongly temperature dependant at cryogenic temperatures and that magnetoresistance has to be taken into account to well assess the Lorentz forces during magnetic field transients. Also, depending of the frequency, skin depth of the materials has to be considered.

### *3.1.2 Mechanical problem formulation*

The mechanical problem is defined and solved from a set of equations (small displacement and small strain assumption):

– Equilibrium equations in static conditions:

- In the body Ω, the relation between the stress tensor and the specific forces is given by:

$$\mathbf{div}\,\underline{\underline{\sigma}} + \boldsymbol{f_v} = \mathbf{0} \text{ in } \Omega \qquad (38)$$

- The boundary conditions with imposed forces reads:

$$\underline{\underline{\sigma}} \cdot \boldsymbol{n} = \boldsymbol{F_s} \text{ on a domain } S_f \qquad (39)$$

– Kinematic:

- Strain tensor is defined by

$$\underline{\underline{\varepsilon}}(\boldsymbol{u}) = \tfrac{1}{2} \cdot \left[\underline{\underline{\mathrm{grad}}}(\boldsymbol{u}) + \underline{\underline{\mathrm{grad}}}^T(\boldsymbol{u})\right] \text{ in } \Omega \qquad (40)$$

- The boundary conditions with imposed displacements reads:

$$\boldsymbol{u} = \boldsymbol{u}_{imposed} \text{ on a domain } S_d \qquad (41)$$

- Constitutive model:
  - The constitutive equation is given by:

$$\underline{\underline{\sigma}} = \underline{\underline{\underline{C}}} : \underline{\underline{\varepsilon}}^e \text{ with } \underline{\underline{\varepsilon}}^e = \underline{\underline{\varepsilon}} - \underline{\underline{\varepsilon}}^{th} - \underline{\underline{\varepsilon}}^p - \ldots \tag{42}$$

For plasticity, it is complemented by plastic flow and consistency equations.
The resolution of the problem leads to the determination of the mechanical solution expressed in term of displacement vector and stress tensor.

### *3.1.3 Design criteria*

#### *3.1.3.1 General rules*

The vacuum chambers have to fulfil different criteria related to material or structural constraints. Typical criteria from a material point of view, are based on the concept of an equivalent stress (Von Mises or Tresca equivalent stress for ductile material; maximum principal stress for brittle material) that shall not exceed an allowed stress limit, usually defined from the material yield stress. Criteria based on fatigue may be required in particular for bellows expansion joints.

From a structural point of view, two main criteria may apply. First, the overall or local deformation is limited. Then, the vacuum chamber stability under external pressure shall be guaranteed. Buckling corresponds to a change of deformation of the structure from compression to bending to minimize its deformation energy.

Whereas material criteria depend on material strength, the buckling criteria are in a first approach independent of the material strength or yield limit and are governed by geometry of the structure and the material stiffness.

#### *3.1.3.2 Structural stability*

For an infinite elastic tube subjected to external pressure, the critical pressure of buckling, $P_{cr}$, is given by:

$$P_{cr} = \frac{E}{4 \cdot (1 - \nu^2)} \cdot \left(\frac{t}{R}\right)^3 \tag{43}$$

A safety factor of 3 is usually applied leading to the design rule of thumb of the minimum thickness of the vacuum chamber in steel given by:

$$t \geq \frac{D}{100} \tag{44}$$

For a beam subjected to a compression axial force, the critical force is given by the Euler's formula:

$$F_{cr} = \left(\frac{\pi}{L_r}\right)^2 \cdot E \cdot I \tag{45}$$

With *I* the area moment inertia and $L_r$, the reduced length determined from the boundary conditions. *I* is approximated by $\pi D^3 t/8$ for a thin tube and is equal to $wh^3/12$ for a rectangular cross-section of height h and width w. The reduced length is shown in Table 6.

**Table 6:** Reduced length for the Euler buckling of beams

| Boundary condition | Reduced length |
|---|---|
| 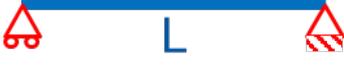 | $L_r = L$ |
| 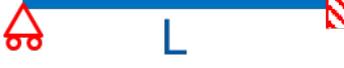 | $L_r = \sqrt{2} \cdot L$ |
| 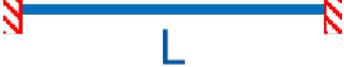 | $L_r = 0.5 \cdot L$ |
| 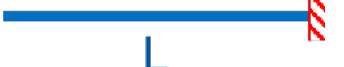 | $L_r = 2 \cdot L$ |

It is worth to mention that the critical buckling load can be significantly influenced by the initial imperfections of the structure.

## 3.2 Vacuum system as a mechanical system

The vacuum system shall be seen as a mechanical system within particular vacuum chambers installed on supports. This assembly shall respect some rules.

### 3.2.1 Supports

#### 3.2.1.1 Kinematic

A vacuum chamber is installed on supports. In the longitudinal direction, only one support has to be fixed. The others have to ensure a free longitudinal movement. This allows to have an isostatic assembly which guaranty a defined position during the installation and stress-free assembly after the installation. In opposite case, large stress may occur during the installation and more over during thermal cycles if applied, leading to the mechanical failure (Eq. (15)).

#### 3.2.1.2 Static

##### 3.2.1.2.1 Deflection

The deflection of a vacuum chamber, of length L and area moment inertia I, simply supported at its extremities is given by:

$$\delta_0 = \frac{5 \cdot q \cdot L^4}{384 \cdot E \cdot I} \quad (46)$$

$E$ is the Young's modulus and $q$ stands for the distributed specific weight (including bake out jackets). This deflection can be reduced adding intermediate support(s) and/or by placing in different locations. For example, for a vacuum chamber simply supported with two supports at the Gaussian points located at around $0.22 \cdot L$ from the extremities, the deflection is reduced by a factor 50.

##### 3.2.1.2.2 Pressure thrust force

For vacuum chambers connected with bellows of different sections, an unbalanced force due to atmospheric pressure is generated and is transferred to the fixed support. This pressure thrust force reads:

$$F_{\to support} = p \cdot \sum_i S_{o_i} \cdot n_i \qquad (47)$$

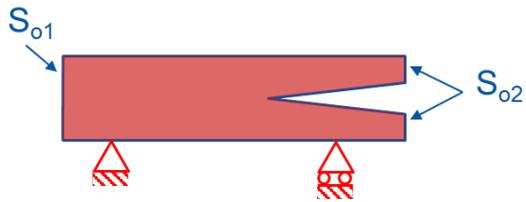

The surface openings, corresponding to bellows, are denoted $S_o$. $n_i$ is the outward normal of the surface. $p$ is the pressure (0.1 MPa).

Large force can be developed, and the solicitations induced by the force on the supports and its anchoring have to be considered carefully.

#### 3.2.1.2.3 Global stability

A pipe, not necessarily with a bellows, may buckle under internal pressure showing a global instability. The buckling pressure can read in a similar way than the Euler's formula. An equivalent length can be defined from the stiffness's and boundary conditions. It governs the buckling mode of the structure (Fig. 10).

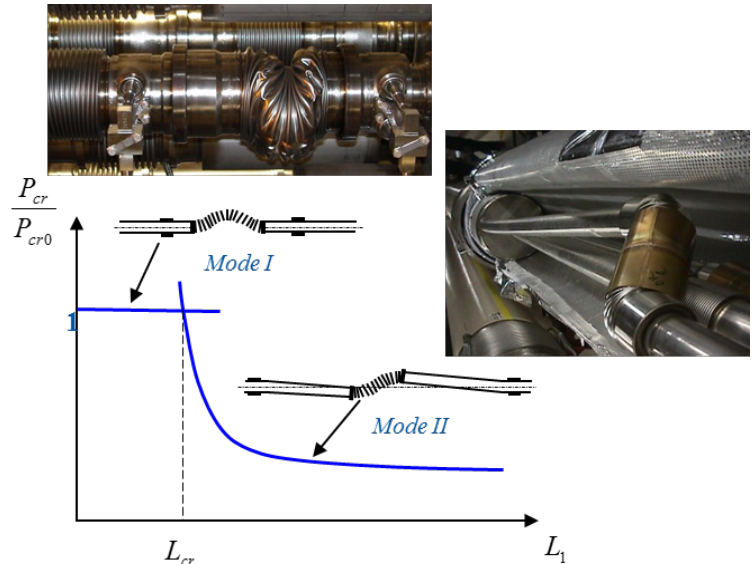

**Fig. 10:** Buckling mode of a line under internal pressure

## 4 Selection criteria of materials

The choice of the materials for the vacuum chamber is usually driven by their main properties. The principal is the vacuum performance (vapour pressure, thermal outgassing, stimulated desorption, SEY, etc.), electrical properties, thermal behaviour, mechanical characteristics (Young's modulus, strength, toughness) and cost.

### 4.1 Figure of merit

For some multi-parameter optimized applications, it is useful to define a figure of merit characterizing the materials with respect to a given criteria. This is the case for example of the vacuum chambers at the interaction points. The chamber has to be as transparent as possible while ensuring the mechanical stability under external pressure. This is translated in having the highest radiation length, thickness ratio with a thickness fulfilling Eq. (43). This leads to a figure of merit given by $X_o \cdot E^{1/3}$.

A few figures of merit are presented below for mechanical stability and beam-material interaction induced heating.

– Mechanical stability for transparent vacuum chamber:
$$X_o \cdot E^{1/3}$$

– Mechanical stability for vacuum chamber subjected to fast magnetic field variation:
$$\rho_e \cdot E^{1/3}$$

– Temperature rise in transient regime:
$$X_o \cdot \rho \cdot C \cdot T_f$$

– Thermal fatigue:
$$\frac{X_o \cdot \rho \cdot C \cdot \sigma_y}{E \cdot \alpha}$$

– Temperature rise in steady state:
$$X_o \cdot \lambda \cdot T_f$$

## 4.2 Material properties

Some material properties of some selected materials used for vacuum applications are summarized in Table 7.

**Table 7:** Indicative properties at room temperature

|  | Unit | Beryllium | Aluminium | Titanium G5 | 316L | Copper | Inconel |
|---|---|---|---|---|---|---|---|
| Density | g·cm$^{-3}$ | 1.85 | 2.8 | 4.4 | 8 | 9 | 8.2 |
| Heat capacity | J·K$^{-1}$·Kg$^{-1}$ | 1830 | 870 | 560 | 500 | 385 | 435 |
| Thermal conductivity | W·K$^{-1}$·m$^{-1}$ | 200 | 217 | 16.7 | 26 | 400 | 11.4 |
| Coefficient of thermal expansion | 10$^{-6}$·K$^{-1}$ | 12 | 22 | 8.9 | 16 | 17 | 13 |
| Radiation length | cm | 35 | 9 | 3.7 | 1.8 | 1.47 | 1.7 |
| Melting temperature | K | 1560 | 930 | 1820 | 1650 | 1360 | 1530 |
| Yield strength | MPa | 345 | 275 | 830 | 300 | 200 | 1100 |
| Young Modulus | GPa | 230 | 73 | 115 | 195 | 115 | 208 |
| Electrical resistivity | 10$^{-9}$·Ω·m | 36 | 28 | 1700 | 750 | 17 | 1250 |

Some properties depend strongly on temperature, grade or delivery state (annealed, hard, …). As examples, the temperature dependences of aluminium alloy strength and copper electrical resistivity are shown in Figs. 11 and 12.

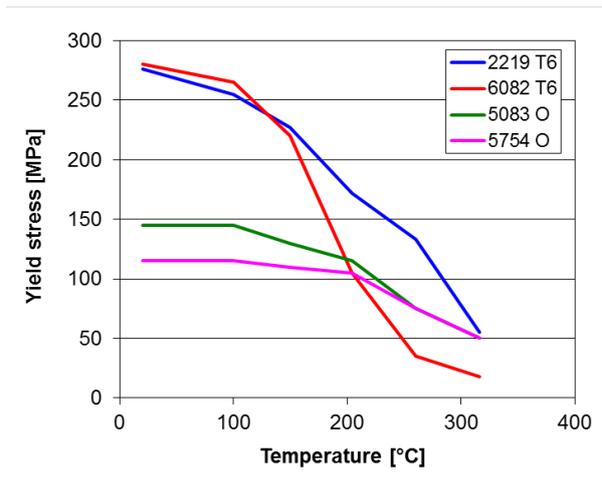
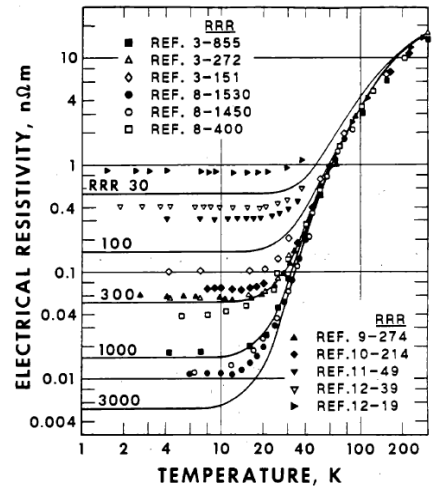

**Fig. 11:** Yield strength of aluminium alloys as a function of temperature

**Fig. 12:** Electrical resistivity of copper as a function of temperature [2]

Some figures of merit of different materials for different criteria are shown in Fig. 13. Values are normalized with respect to beryllium.

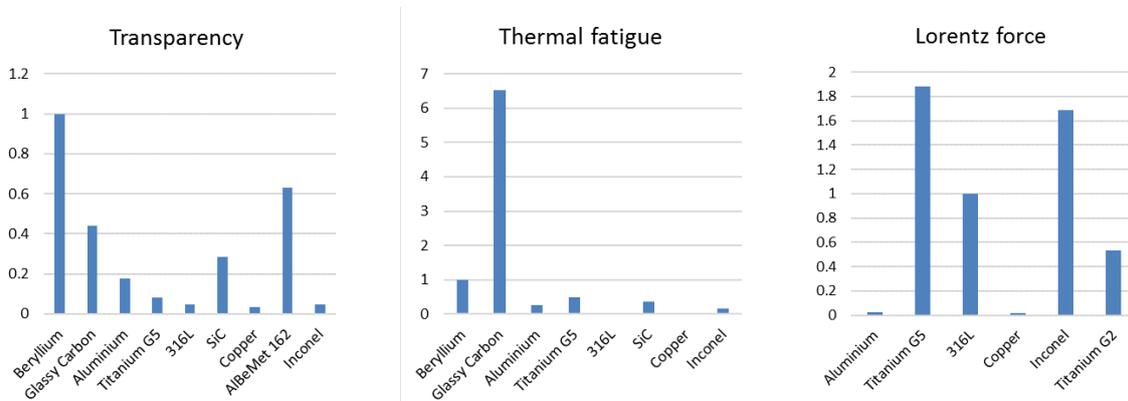

**Fig. 13:** Figures of merit

Difference between materials is significant and therefore the choice of the appropriate material can be crucial in some applications.

**Summary**

For most of vacuum systems, the mechanical design study can be treated in the framework of the thermal elasticity. The corresponding properties, in particular thermal expansion coefficient and the material stiffness (Young's modulus) governs the structural behaviour of a vacuum chamber: thermal dilatation/contraction, deformation and stability under vacuum. For advanced designs, additional approaches and models might be necessary (plasticity, damage and fracture mechanics) and some hints have been given in the paper.

Some design rules of a vacuum system have been given. A few have to be kept in mind of all mechanical vacuum engineer: First, the stability against the buckling under external pressure; second, the vacuum thrust force induced by bellows with different cross-sections; and third, the use of only one longitudinal fixed support. Applying these simple rules in a systematic approach will help to get reliable design.